\documentclass[aps,twocolumn,showpacs,groupedaddress,nofootinbib]{revtex4-1}  % for review and submission 
\usepackage{graphicx}  % needed for figures
\usepackage{dcolumn}   % needed for some tables
\usepackage{bm}        % for math
\usepackage{amssymb}   % for math
\usepackage{amsmath}
\usepackage{epsfig}
\usepackage{xspace}

\newcommand{\epsf}{\epsilon^{1/4}}

\renewcommand{\vec}[1]{{\bf #1}}

\begin{document}

% the following line is for submission, including submission to the arXiv!!
%\hspace{5.2in} \mbox{Darmstadt/DESY/U-Tokyo}

\title{Out of equilibrium dynamics of coherent non-abelian gauge fields}
\author{J{\"u}rgen Berges$^{1,3}$}
\author{Sebastian Scheffler$^{2}$}
\author{S\"oren Schlichting$^{1,2}$}
\author{D\'enes Sexty$^{1,3}$}
\affiliation{$^{1}$Institute for Theoretical Physics, Heidelberg University, Philosophenweg 16, 69120 Heidelberg}
\affiliation{$^{2}$Institute for Nuclear Physics, Darmstadt University of Technology, Schlossgartenstr.\ 9, 64289 Darmstadt, Germany}
\affiliation{$^{3}$ExtreMe Matter Institute (EMMI),
             GSI Helmholtzzentrum f\"ur Schwerionenforschung GmbH, 
             Planckstra\ss e~1, 
             64291~Darmstadt, Germany}

\begin{abstract} 
We study out-of-equilibrium dynamics of intense non-abelian gauge fields. Generalizing the well-known Nielsen-Olesen instabilities for constant initial color-magnetic fields, we investigate the impact of temporal modulations and fluctuations in the initial conditions. This leads to a remarkable coexistence of the original Nielsen-Olesen instability and the subdominant phenomenon of parametric resonance. Taking into account that the fields may be correlated only over a limited transverse size, we model characteristic aspects of the dynamics of color flux tubes relevant in the context of heavy-ion collisions.
\end{abstract}

\maketitle

\section{Introduction and overview}\label{sec:introduction}

Out of equilibrium dynamics of non-abelian gauge fields provides an outstanding challenge in quantum field theory. Its most prominent application concerns the description of relativistic heavy ion collisions. It involves the solution of an initial value problem in real time based on the underlying theory of quantum chromodynamics (QCD). This is not accessible to standard lattice gauge theory simulations in Euclidean space-time. However, important aspects of nonequilibrium gauge field dynamics may be described using classical-statistical lattice gauge theory techniques in real time. Classical-statistical approximations to the quantum dynamics are expected to be valid if the expectation value of the anti-commutator of fields is much larger than the commutator. Loosely speaking this is realized in the presence of sufficiently high occupation numbers per mode or large coherent field expectation values. 

The physics of nonequilibrium instabilities provides an important
example where classical-statistical simulations techniques can give
accurate descriptions of the underlying quantum dynamics. This has
been tested to high accuracy for self-interacting scalar quantum field
theories~\cite{Aarts:2001yn,Arrizabalaga:2004iw,Berges:2008wm} and Yukawa-like
theories with fermions~\cite{Berges:2010zv}, where appropriate
far-from-equilibrium resummation techniques are available for the
respective quantum field theory. In recent years classical simulation
techniques have become an important building block for our
understanding of the time evolution of instabilities in non-abelian
gauge theories in the context of collision experiments with heavy
nuclei~\cite{Romatschke:2005pm,Berges:2007re,Kunihiro:2010tg,Fukushima:2011nq}.

The present work investigates classical dynamics of coherent
non-abelian gauge fields, which is also motivated by the notion of
'color flux tubes' that may form after the collision of two Lorentz
contracted heavy nuclei. These are characterized by intense
color-magnetic as well as color-electric field configurations in
longitudinal direction. They are smooth in this direction and
correlated over a transverse size associated to the inverse of the
characteristic momentum scale $Q_s$ in the saturation
scenario~\cite{Gelis:2010nm}. To understand the time evolution of this
configuration in QCD is a formidable task, which is further
complicated by the longitudinal expansion of the system.

In order to approach this complex question of non-linear gauge field dynamics, it is very instructive to consider first an extreme simplification which can also give analytical insights. For the ease of a later numerical treatment,
we use $SU(2)$ gauge theory. We neglect expansion and employ a constant color-magnetic field configuration, $B^a_\mu$ with color index $a=1,2,3$ and Lorentz index $\mu$, whose spatial components $i=x,y,z$ point in longitudinal ($i=z$) direction  
\begin{equation}
B_i^a \ = \, \delta^{1 a} \delta_{z i} B \, .
\label{eq:Blong}
\end{equation}
Here the latter can always be arranged to point into the direction $a = 1$ in color
space by a suitable gauge transformation without loss of generality. In the context of heavy ion collisions this configuration has been extensively discussed in Ref.~\cite{Iwazaki:2008xi}, where the field in infinite volume is taken to be generated from the vector potential $A^a_\mu$ as
\begin{equation}
A_x^1 = - \frac{1}{2} y B \, , \quad A_y^1 = \frac{1}{2} x B \, , 
\label{eq:NO-A}
\end{equation}
with all other spatial components vanishing.
Such a field configuration is known to exhibit a Nielsen-Olesen instability~\cite{NO} characterized by an exponential growth of fluctuations with maximum growth rate 
\begin{equation}
\gamma_{\rm NO} \, = \, \sqrt{g B} \, ,
\label{eq:NOmaxrate}
\end{equation}
where $\sqrt{g B}$ may be taken to be of order $Q_s$. This exponential growth leads to a production of gluons, which is much faster than any conventional production process. Its consequences for the question of thermalization in a heavy ion collision can be significant. Of course, in this context a single macroscopic constant color-magnetic background field is certainly not a realistic option and it is important to understand the robustness of the underlying physical processes against suitable generalizations.     

In a first generalization we will allow for temporal modulations of the color magnetic background field configurations. This can be conveniently achieved by the gauge field configuration 
\begin{equation}
 A_x^2 \, = \, A_y^3 \, = \, \sqrt{\frac{B}{g}} \, , 
\label{eq:Anonlinear}
\end{equation}
with all other components zero. It can be directly verified that this vector potential configuration gives the same form of the longitudinal magnetic field (\ref{eq:Blong}). However, there are important differences as compared to the previously considered case, where the magnetic field is generated from space-dependent vector potentials. In contrast to (\ref{eq:NO-A}), the configuration (\ref{eq:Anonlinear}) contributes to the non-linear part of the field strength tensor
\begin{equation}
F^a_{\mu\nu}[A] \, = \, \partial_\mu A^a_\nu - \partial_\nu A^a_\mu + g \epsilon^{abc} A^b_\mu A^c_\nu
\label{eq:fieldstrength}
\end{equation}
from which the magnetic field is obtained as $B^{a}_j = \epsilon^{ijk} F^a_{jk}$. Generating the magnetic field with (\ref{eq:Anonlinear}) via the non-linear term of the field strength tensor leads to coherent oscillations in time with characteristic frequency $\sim \sqrt{g B}$. In contrast, it can be readily verified that (\ref{eq:NO-A}) is a time-independent solution of the
classical Yang-Mills equations. Accordingly, also the
magnetic field derived from~(\ref{eq:NO-A}) is constant in
both space and time as long as it is unperturbed. 

We discuss analytic solutions for the classical time evolution of the configuration (\ref{eq:Anonlinear}) in temporal (Weyl) gauge and study the behavior of linear perturbations on top of the oscillating background field in Sec.~\ref{sec:linear}. It is shown that these perturbations still exhibit robust growth similar to the Nielsen-Olesen--type. In order to go beyond the linear analysis we employ classical-statistical lattice gauge theory simulations in Sec.~\ref{sec:classical-statistical}. The linear analysis is seen to accurately reproduce the numerical data for times which are short compared to the inverse characteristic growth rates. Afterwards, non-linear contributions become crucial. An important observation is that the non-linearities lead to a very efficient growth for all components of the gauge potential such that the details about the initial configuration (\ref{eq:Blong}) become irrelevant as long as sufficient primary growth is triggered. This adds to the robustness of the observed phenomena also for further generalizations of the initial configurations.

An important step towards more realistic scenarios concerns the inclusion of fluctuations into the initial conditions. The initial configuration (\ref{eq:Anonlinear}) exhibits a sharply defined macroscopic field amplitude $B$ which is certainly unrealistic, in particular, since no macroscopic colored objects can exist. Here the time-evolution of single macroscopic $B$-field configurations may be viewed as describing the dynamics of individual members of an ensemble. Ensemble averages are then obtained from sampling initial conditions according to given averages or expectation values at initial time. In Sec.~\ref{sec:coherent} we consider nonequilibrium ensembles, where we choose the initial values of the spatially homogeneous fields in (\ref{eq:Anonlinear}) randomly from a Gaussian distribution with zero mean. Again one finds robust growth also for the ensemble of homogeneous initial fields. A dramatic difference is that for the ensemble average one observes isotropization of the stress-energy tensor happening on a time scale much faster than $1/\sqrt{gB}$ because of a dephasing phenomenon. We comment on the possible relevance of this observation for the outstanding question of rapid isotropization in heavy ion collisions in Sec.~\ref{sec:iso}.

Finally, we take into account that the fields should be correlated only over a limited transverse size, which in the color flux tube picture is associated to the inverse of the characteristic momentum scale $Q_s$. For this we divide in Sec.~\ref{sec:patches} the transverse coordinate plane into domains of equal spatial size. Each domain is then filled with a coherent color magnetic field, however, using a different random color direction in each patch. If the transverse size is taken to be $Q_s$ or $\sqrt{gB}$ then this should model characteristic aspects of the dynamics of color flux tubes. It turns out that one observes exponential growth for all components of this inhomogeneous gauge field configuration. The nature of the underlying instability still shows properties reminiscent of the Nielsen-Olesen--type. In particular, the characteristic dependence of the primary growth rate as a function of momentum still reaches its maximum at low momenta.
In contrast, Weibel instabilities show a vanishing growth rate at zero momentum~\cite{Weibel}. 
This difference has also been emphasized in Ref.~\cite{Iwazaki:2008xi}. We conclude in Sec.~\ref{sec:conclude}. In an appendix we describe that the phenomenon of parametric resonance leads to a subdominant instability band at higher longitudinal momenta.

\section{Linear regime}
\label{sec:linear}

We first discuss analytic solutions for the classical time evolution starting from the coherent field configuration (\ref{eq:Anonlinear}). It is shown that linear perturbations on top of the oscillating background field still exhibit robust growth similar to the Nielsen-Olesen--type.
We consider a non-abelian gauge theory with $SU(2)$ color gauge group.
Taking into account two colors simplifies the analysis as compared to the $SU(3)$ gauge group relevant for QCD and the difference is expected to be of minor relevance for the physics considered here \cite{Berges:2008zt,Ipp:2010uy}. With the covariant derivative
\begin{equation}
D_\mu^{ab}[A] \, = \, \partial_\mu \delta^{ab} + g \epsilon^{acb} A_\mu^c
\end{equation}
the classical equations of motion read
\begin{equation}
\left(D_\mu[A] F^{\mu\nu}[A]\right)^a \, = \, 0 \, .
\label{eq:classicalfieldequation}
\end{equation}

In a first step, we will compute the classical time evolution starting from the coherent field configuration (\ref{eq:Anonlinear}). For an analytical understanding of the dynamics it is convenient to split the gauge field potentials $A^a_\mu(x)$ into a
time-dependent background field $\bar{A}^a_\mu(x^0)$ and a perturbation:
\begin{equation}
A^a_\mu(x) \, = \, \bar{A}^a_\mu(x^0) + \delta A^a_\mu(x) \, .
\label{eq:background}
\end{equation}
We can then compute the time evolution in an expansion in powers of the perturbation $\delta A^a_\mu(x)$. At zeroth order we obtain the field equation for the background field 
\begin{equation}
\left(D_\mu[\bar{A}] F^{\mu\nu}[\bar{A}]\right)^a \, = \, 0 \, .
\label{eq:backgroundfieldequation}
\end{equation}
The next order corresponds to the linearized equation for the
fluctuations \cite{Tudron:1980gq}, 
\begin{eqnarray}
&&\left(D_\mu[\bar{A}]D^\mu[\bar{A}]\delta A^\nu\right)^a - \left(D_\mu[\bar{A}]D^\nu[\bar{A}]\delta A^\mu\right)^a  \nonumber\\
&& + g \epsilon^{abc} \delta A^b_\mu F^{c\mu\nu}[\bar{A}] \, = \, 0 \, .
\label{eq:fluctuationequation}
\end{eqnarray}
Accordingly, equations (\ref{eq:backgroundfieldequation}) and (\ref{eq:fluctuationequation}) correspond to the classical field equation (\ref{eq:classicalfieldequation}) up to corrections of order $(\delta A)^2$.

\begin{figure}
\begin{center}
 \epsfig{file=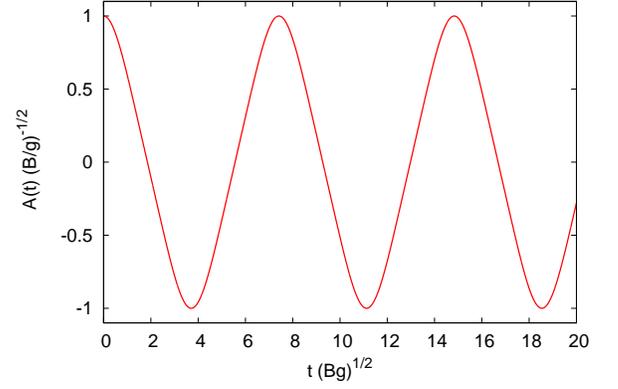, width=8cm} 
 \caption{Time evolution of the background field $\bar{A}(t)$ normalized to its initial value. 
It is described by the Jacobi elliptic function 
${\rm cn}(\sqrt{Bg}\, t,1/2)\,\sqrt{B/g}$ as a function of time $t$ in units of the characteristic scale $\sqrt{gB}$.}\label{fig:jacobiancn}
\end{center}
\end{figure}

We choose temporal (Weyl) gauge where $A^a_0 = 0$. Writing $t \equiv x^0$ we consider the field configuration 
\begin{equation}
\bar{A}^a_i(t) \, = \, \bar{A}(t) \left( \delta^{a2} \delta_{ix} + \delta^{a3} \delta_{iy} \right)
\end{equation}
which corresponds to (\ref{eq:Anonlinear}) at initial time with $\bar{A}(t=0) = \sqrt{B/g}$ for given initial color-magnetic field $B$. With the further initial condition $\partial_t\bar{A}(t=0) = 0$ the background field equation (\ref{eq:backgroundfieldequation}) reads
\begin{equation}
\partial_t^2 \bar{A}(t) + g^2 \bar{A}(t)^3 \, = \, 0
\end{equation}
and the solution is given in terms of a Jacobi elliptic function:
\begin{equation}
\bar{A}(t) \, = \, \sqrt{\frac{B}{g}} \,\, {\rm cn}\!\left(\sqrt{g B}\, t\,,\frac{1}{2}\right) \, .
\label{eq:Asol}
\end{equation}
Here ${\rm cn}(\sqrt{g B}\, t,1/2)$ is an oscillatory function such that $\bar{A}(t) = \bar{A}(t+\Delta t_B)$ with period 
\begin{equation}
\Delta t_B \, = \, \frac{4 K(1/2)}{\sqrt{g B}} \,\simeq \, \frac{7.42}{\sqrt{g B}} \, ,
\end{equation}
where $K(1/2)$ denotes the complete elliptic integral of the first kind \cite{Elliptic}. Similar behavior 
has been discussed in the context of classical chaos \cite{chaos}, or for parametric resonance in scalar field theories \cite{Kofman:1994rk}. The solution (\ref{eq:Asol}) is plotted in Fig.~\ref{fig:jacobiancn}. For later use we note that the corresponding characteristic frequency is 
$\omega_B = 2 \pi/\Delta t_B \simeq 0.847 \sqrt{g B}$. We emphasize that 
no constant non-zero solution exists for $\bar{A}$ in this case. This is in contrast to the analysis of Ref.~\cite{Tudron:1980gq}, where the same gauge potential configuration in the presence of suitable external charges is assumed to lead to a time-independent $\bar{A}$.

In the following we insert this background field behavior into the linearized fluctuation equation (\ref{eq:fluctuationequation}) and discuss solutions for modes of $\delta A^a_i$ in spatial Fourier space:
\begin{equation}
\delta A^a_i(t,{\bf p}) \, = \, \int {\rm d}^3 x\, e^{-i p_j x^j} \delta A^a_i(t,{\bf x})
\end{equation}
We concentrate on momenta $p_z$ along the $z$-direction, i.e.\ we evaluate the modes of $\delta A^a_i(t,{\bf p})$ at $p_x=p_y=0$. The evolution equation (\ref{eq:fluctuationequation}) for the different components of $\delta A^a_i(t,p_z)$ can then be written with $ \delta \dot{A}^a_i(t,p_z) \equiv \partial_t \delta A^a_i(t,p_z)$ as the following independent sets of coupled differential equations: 
\begin{equation} 
\left( \begin{array}{c}
\delta \ddot{A}^1_x \\
\delta \ddot{A}^3_z \end{array} \right) \, = \, 
-\left( \begin{array}{cc}
g^2 \bar{A}^2 + p_z^2 & i g \bar{A} p_z  \\
-i g \bar{A} p_z & g^2 \bar{A}^2  \end{array} \right)
\left( \begin{array}{c}
\delta A^1_x \\
\delta A^3_z \end{array} \right) .
\label{eq:M1}
\end{equation}
The complex conjugate of the components $(\delta A^1_y, \delta A^2_z)$ obey the same equation, which is obtained from (\ref{eq:M1}) with the replacement $(\delta A^1_x, \delta A^3_z) \, \rightarrow \, (\delta A^{1*}_y, \delta A^{2*}_z)$. Similarly, one finds
\begin{equation} 
\left( \begin{array}{c}
\delta \ddot{A}^2_x \\
\delta \ddot{A}^3_y \end{array} \right) \, = \, 
-\left( \begin{array}{cc}
g^2 \bar{A}^2 + p_z^2 & 2 g^2 \bar{A}^2 \\
2 g^2 \bar{A}^2 & g^2 \bar{A}^2 + p_z^2 \end{array} \right)
\left( \begin{array}{c}
\delta A^2_x \\
\delta A^3_y \end{array} \right)
\label{eq:M2}
\end{equation}
and
\begin{equation} 
\left( \begin{array}{c}
\delta \ddot{A}^3_x \\
\delta \ddot{A}^2_y \\
\delta \ddot{A}^1_z \end{array} \right) \, = \, 
-\left( \begin{array}{ccc}
p_z^2 & -g^2 \bar{A}^2 & -i g \bar{A} p_z \\
-g^2 \bar{A}^2 & p_z^2 & i g \bar{A} p_z \\
i g \bar{A} p_z & -i g \bar{A} p_z & 2 g^2 \bar{A}^2 \end{array} \right)
\left( \begin{array}{c}
\delta A^3_x \\
\delta A^2_y \\
\delta A^1_z \end{array} \right).
\label{eq:M3}
\end{equation} 
Each of the linear differential equation matrices (\ref{eq:M1})-(\ref{eq:M3}) with time dependent background field $\bar{A}(t)$ may be further analyzed by diagonalization. The time-dependent eigenvalues of the equation matrices read for (\ref{eq:M1})
\begin{eqnarray}
\omega_{1}^{\pm}(p_z)^2 &=& g^2 \bar{A}^2  + \frac{p_z^2}{2} \pm \frac{1}{2} \sqrt{4 g^2 \bar{A}^2 p_z^2 + p_z^4} \, , 
\end{eqnarray}
for (\ref{eq:M2})
\begin{eqnarray}
\omega_2(p_z)^2 &=& 3 g^2 \bar{A}^2 + p_z^2 \, , \label{eq:omega}\\
\omega_3(p_z)^2 &=& - (g^2 \bar{A}^2 - p_z^2) \, , \nonumber
\end{eqnarray}
and for (\ref{eq:M3})
\begin{eqnarray}
\omega_4(p_z)^2 &=& - (g^2 \bar{A}^2 - p_z^2) \, , \\ 
\omega_{5}^{\pm}(p_z)^2 &=& \frac{1}{2} \left(3 g^2 \bar{A}^2 + p_z^2 \pm \sqrt{
    g^4 \bar{A}^4  + 6 g^2 \bar{A}^2 p_z^2 + p_z^4}\right) \, . \nonumber
\end{eqnarray}
The corresponding eigenvectors depend, in general, on $\bar{A}(t)$ and thus on time. However, the eigenvectors associated to $\omega_2(p_z)^2$, $\omega_3(p_z)^2$ and $\omega_4(p_z)^2$ are time-independent and given by $(1,1)/\sqrt{2}$, $(-1,1)/\sqrt{2}$, and $(1,1,0)/\sqrt{2}$, respectively. These include, in particular, the only {\em negative} eigenvalues $\omega_3(p_z)^2 = \omega_4(p_z)^2$ for momenta $p_z^2 < g^2 \bar{A}(t)$. These will play a particularly important role in the following since they turn out to govern the fastest time scales.

It is very instructive to consider the example of (\ref{eq:M2}) in diagonalized form, i.e.\
\begin{eqnarray}
\delta \ddot{A}_+ &=& -\left(3 g^2 \bar{A}^2 + p_z^2 \right) \delta A_+  \, , 
\label{eq:Ap}\\
\delta \ddot{A}_- &=& \left(g^2 \bar{A}^2 - p_z^2 \right) \delta A_- \, , \nonumber 
\end{eqnarray}
where $\delta A_+ = \delta A^2_x + \delta A^3_y$ and $\delta A_- = \delta A^3_y - \delta A^2_x$. (The same equation for $\delta A_-$ would also be obtained from (\ref{eq:M3}) by associating $\delta A_-$ to $\delta A^3_x + \delta A^2_y$.) These equations are of the Lam{\'e}-type \cite{Elliptic}: Since the squared background field appearing in (\ref{eq:Ap}) has periodicity $\Delta t_B/2$, each solution can be written as a linear combination of the form
\begin{equation}
\delta A_+(t+\Delta t/2,p_z) = e^{i C_+(p_z)}\, \delta A_+(t,p_z) 
\end{equation}
such that 
\begin{equation} \label{eq:linansatz}
\delta A_+(t,p_z) = e^{2i C_+(p_z) t/\Delta t_B}\, \Pi_+(t,p_z) 
\end{equation}
with periodic functions $\Pi_+(t+\Delta t/2,p_z) = \Pi_+(t,p_z)$ and similarly for $\delta A_-(t,p_z)$. The Floquet function $C_+(p_z)$ is time-independent and leads to oscillating behavior if $C_+(p_z)$ is real or to exponentially growing or decaying solutions if imaginary.  

The dominant exponentially growing solutions arise because of the appearance of the time-dependent negative eigenvalues $\omega_3(p_z)^2=\omega_4(p_z)^2$ and we will concentrate on them in the following. Subdominantly growing modes are associated to parametric resonance and are discussed in detail in the appendix. We first observe from the evolution equation (\ref{eq:Ap}) for $\delta A_-(t,p_z)$ that replacing the background field by a constant, $\bar{A}(t) \rightarrow \sqrt{B(t=0)/g}$, would lead to the well-known Nielsen-Olesen result for the growth rate
of modes with $p_z^2 \le g B(t=0)$:
\begin{equation} 
\gamma_{\rm NO}(p_z) \, = \, \sqrt{g B(t=0) - p_z^2} \, .  
\label{eq:gammano}
\end{equation}
This is in accordance with the fact that the value for the maximum growth rate for constant fields, as stated in (\ref{eq:NOmaxrate}), is obtained for vanishing momenta. In contrast, since the background field is oscillatory in our case there will be deviations from the Nielson-Olesen result (\ref{eq:gammano}). However, it turns out that to rather good accuracy their growth rates follow the Nielsen-Olesen estimate if the temporal average of the oscillating magnetic background field is used, which is explained in the appendix. The time average over one period $\Delta t_B/2$ of the square of the time-dependent background field (\ref{eq:Asol}), which enters the evolution equation for the fluctuation $\delta A_-(t,p_z)$, is given for the above initial conditions by 
\begin{eqnarray}
g \overline{B} & \equiv & \frac{g B(t=0)}{2 K(1/2)} \int_0^{2 K(1/2)} \! {\rm d}x \, {\rm cn}^2\!\left(x, \frac{1}{2}\right) \\
& = & \frac{ \Gamma(3/4)^2 }{ \Gamma(5/4) \Gamma(1/4) }\,  g B(t=0) \nonumber\\
& \approx & 0.457\, g B(t=0) \, . \label{eq:averageB}
\end{eqnarray}
Indeed, replacing $g B(t=0)$ in (\ref{eq:gammano}) by the average value (\ref{eq:averageB}) reproduces the full numerical results in the linear regime at sufficiently early times to good accuracy, which is described in the following.

\section{Classical lattice gauge theory}
\label{sec:classical-statistical}

The exponentially growing solutions observed in the previous section will finally lead to non-linear behavior. In order to be able to describe this physics, we use classical-statistical Yang-Mills theory in Minkowski space-time following closely Ref.~\cite{Berges:2007re}, to which we refer for further technical details. The above linear analysis will be seen to accurately reproduce the numerical data for times which are short compared to the inverse characteristic 'primary' growth rates of the linear theory. Afterwards, non-linear contributions become crucial. In particular, the primary growth induces 'secondary' growth rates which are multiples of the primary ones similar to what has been observed in Refs.~\cite{Berges:2007re}, where no coherent magnetic fields were employed. As a consequence, a very efficient growth for all components of the gauge potential is observed such that the details about the initial configuration (\ref{eq:Blong}) become irrelevant as long as sufficient primary growth is triggered. 

For the simulations we employ the Wilsonian lattice action for SU($2$) gauge theory in Minkowski space-time:
\begin{eqnarray}
S[U] &=& - \beta_0 \sum_{x} \sum_i \left\{ \frac{1}{2 {\rm Tr}
\mathbf{1}} \left( {\rm Tr}\, U_{x,0i} + {\rm Tr}\, U_{x,0i}^{\dagger}
\right) - 1 \right\}
\nonumber\\
&& + \beta_s \sum_{x} \sum_{i<j} \left\{ \frac{1}{2 {\rm
Tr} \mathbf{1}} \left( {\rm Tr}\, U_{x,ij} + {\rm Tr}\, U_{x,ij}^{\dagger}
\right) - 1 \right\} ,
\nonumber\\
\label{eq:LatticeAction}
\end{eqnarray}
with $x = (x^0, {\bf x})$ and spatial Lorentz indices $i,j = 1,2,3$. It is given in terms of the plaquette variable $U_{x,\mu\nu} \equiv U_{x,\mu} U_{x+\hat\mu,\nu}
U^{\dagger}_{x+\hat\nu,\mu} U^{\dagger}_{x,\nu}$,
where $U_{x,\nu\mu}^{\dagger}=U_{x,\mu\nu}\,$. Here $U_{x,\mu}$ is the
parallel transporter associated with the link from the neighboring
lattice point $x+\hat{\mu}$ to the point $x$ in the direction of
the lattice axis $\mu = 0,1,2,3$. The definitions
$\beta_0 \equiv 2 \gamma {\rm Tr} \mathbf{1}/g_0^2$ and
$\beta_s \equiv 2 {\rm Tr} \mathbf{1}/(g_s^2 \gamma)$
contain the lattice parameter $\gamma \equiv a_s/a_t$, where $a_s$ denotes the spatial and $a_t$ the temporal lattice spacings, and we will consider $g_0 = g_s = g$. The dynamics is solved in temporal axial gauge. 

Varying the action (\ref{eq:LatticeAction}) w.r.t.\ the spatial link variables $U_{x, j}$ yields the classical lattice equations of motion. Variation w.r.t.\ to a temporal link gives the Gauss constraint. The coupling $g$ can be scaled out of the classical equations of motion and we set $g = 1$ for the simulations. We define the gauge fields as
\begin{equation}
g A_i^a(x) \,=\, -\frac{i}{2 a_s} \, {\rm Tr} \left( \sigma^a U_i(x) \right) 
\label{eq:compute-gauge-field}
\end{equation}
where $\sigma^1$, $\sigma^2$ and $\sigma^3$ denote the three Pauli matrices. Correlation functions are obtained by repeated numerical integration of the classical lattice equations of motion and Monte Carlo sampling of initial conditions~\cite{Berges:2007re}.  The initial time derivatives $\dot{A}_{\mu}^a(t=0, \vec{x})$ are set to zero for all shown results, which implements the Gauss constraint at all times. This simplifies the numerical implementation, however, we checked that including initial electric fields does not change much the observed growth rates. Shown results are from computations on $N^3 = 128^3$ lattices. 

\begin{figure}
\begin{center}
 \epsfig{file=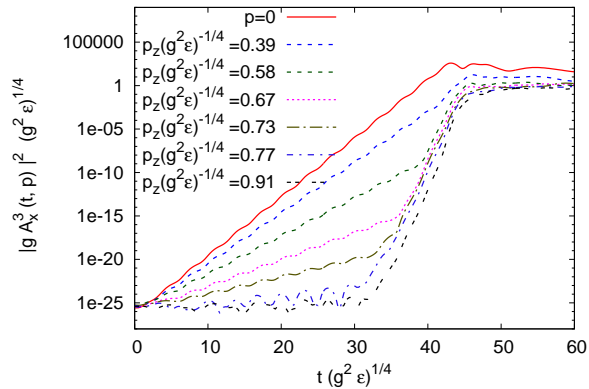, width=8cm}
 \caption{Time evolution of spatial Fourier modes of $A_x^3$ as a function of time, all in appropriate units of the energy density $\epsilon$. The momenta are chosen in the $z$ direction.}
\label{fig1}
\end{center}
\end{figure}
Using the initial condition (\ref{eq:Anonlinear}), supplemented by 
a small noise for all gauge field components, we indeed find exponentially growing behavior corresponding to a nonequilibrium instability. In Fig.~\ref{fig1} we show the absolute value squared of $A^3_x(t,{\bf p})$ as a function of time in units of the fourth root of the energy density $\epsilon$. Different lines correspond to different spatial momenta and one observes that modes with smallest momentum have the biggest initial 
growth rate, as expected from the linear analysis of Sec.~\ref{sec:linear}. In contrast to the linear analysis, Fig.~\ref{fig1} exhibits that higher momentum modes can also have significant growth at somewhat later times.  

\begin{figure}
\begin{center}
\epsfig{file=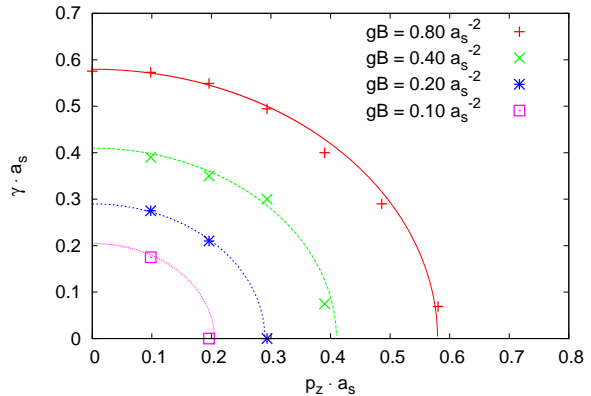, width=8.0cm, angle=0}
\hspace{1cm}
\epsfig{file=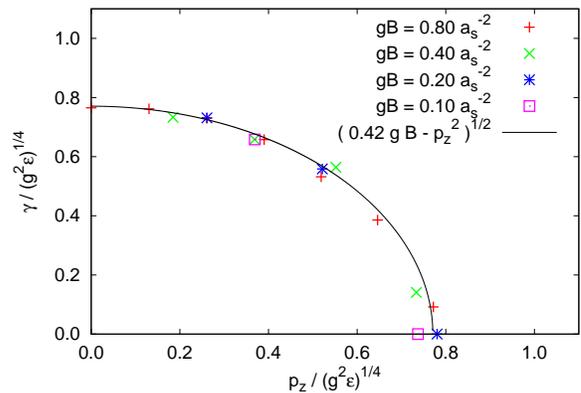, width=8.0cm, angle=0}
\caption{Growth rates for $|A_x^3(t,p_z)|$ from simulations
  with different magnetic field strengths measured (a) in lattice
  units and (b) in units of $\epsf $. The solid lines represent the
  function (\ref{eq-NO-growth-rates-num}) for the
  various magnetic field strengths. The presence of a subdominant
instability band is shown in the appendix, in Fig.\ref{fig-anares}.
 }
\label{fig-NO-growth-rates}
\end{center}
\end{figure}
Concentrating first on earlier times, we obtain an average growth rate 
by fitting an exponential function to  $|A_x^3(t,p_z)|$
 over times large
compared to the oscillation frequency. The results
of such a fit are shown in Fig.~\ref{fig-NO-growth-rates}. The 
upper panel shows the growth rate as a function of $p_z$ for different
initial magnetic field strengths.  Both the growth rate and the
momentum are plotted in lattice units. 
We find that the primary growth
rates can be described very well by
\begin{equation}
\gamma^{(\text{num})}(p_z) \, \simeq \, \sqrt{0.42 g B(t=0) - p_z^2} 
\label{eq-NO-growth-rates-num}
\end{equation}
for momenta $p_z^2 \lesssim 0.42 g B(t=0)$. This function is displayed as a curve in
Fig.~\ref{fig-NO-growth-rates} (a) along with the fit values. The numerical 
factor appearing in front of $g B(t=0)$ in 
(\ref{eq-NO-growth-rates-num}) is equal within errors to the one
in (\ref{eq:averageB}) found from the linear analysis. 
\begin{figure*}[!ht]
\begin{center}
 \epsfig{file=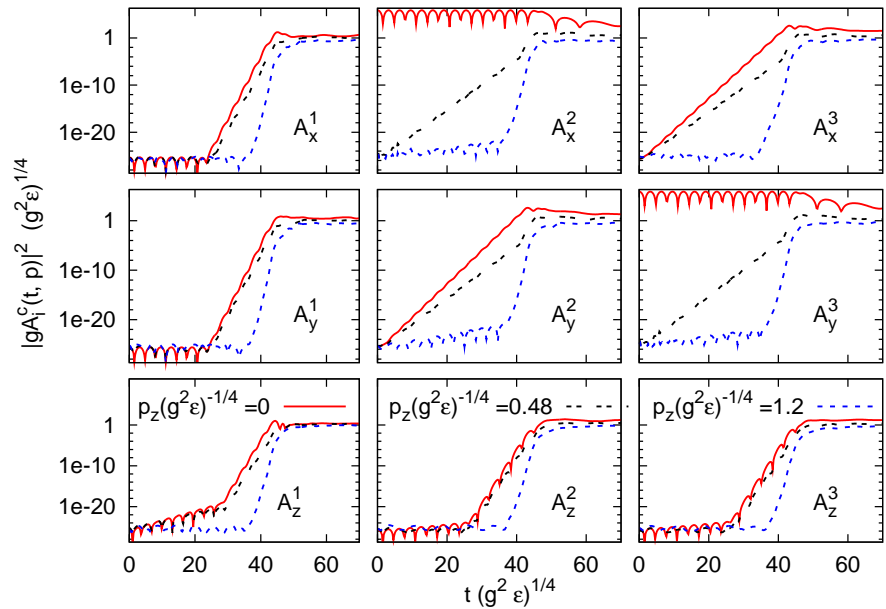, width=12cm}
\caption{Time evolution of the individual gauge field components in
   units of appropriate powers of the energy density. Each panel
   shows the time evolution of three different Fourier coefficients
   whose momenta are parallel to the $z$-axis.}
\label{fig:NO-A-evolution}
\end{center}
\end{figure*}

In the lower panel of Fig.~\ref{fig-NO-growth-rates} (b) we plot the same data as in (a), however, this time the growth
rates as well as momenta are measured in units of $ (g^2\epsilon)^{1/4}$. One observes a collapse of the results from simulations with different
initial field strength onto a single curve to very good accuracy.
The reason for this trivial scaling can be understood from
(\ref{eq-NO-growth-rates-num}) for
the growth rate and the fact that the energy density is proportional to $B^2$. 
We note that parametric resonance leads to a subdominant instability band at higher 
longitudinal momenta, which is given in the appendix.

Fig.~\ref{fig1} shows that there are significant deviations from 
the linear analysis for higher momentum modes already at rather early times, before the growth of the lowest momentum mode saturates. Though these higher momentum modes have initially comparably small primary growth rates, there is a substantial speed-up because of non-linearities at later times. As a consequence, they almost catch up with the initially fastest growing mode. The observed non-linear behavior is the consequence of the self-interactions of the gauge fields \cite{Berges:2007re}.
The secondary growth rate is to a good approximation three times the growth 
rate of the fastest primary growth, which can be described 
from resummed loop expansions based on the two-particle irreducible 
effective action following the lines of Ref.~\cite{Berges:2002cz}, where similar
phenomena for nonequilibrium instabilities in scalar field theories
were studied. 

\begin{figure}
\begin{center}
 \epsfig{file=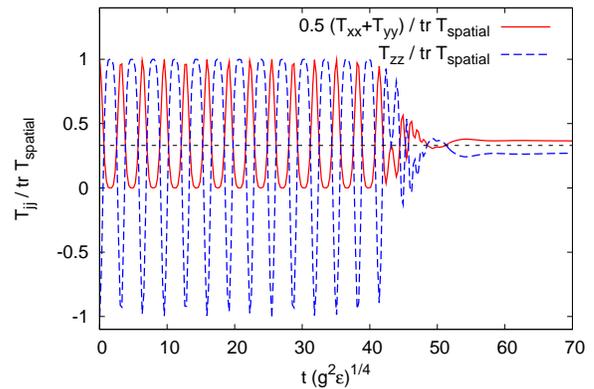, width=8cm}
 \caption{Time evolution of the diagonal entries of the spatial components of the
   stress-energy tensor, i.e.\ transverse versus longitudinal 'pressure', for single runs with a macroscopic initial $B$-field configuration defined in (\ref{eq:Anonlinear}). }
\label{fig-NO-T_ij} 
\end{center}
\end{figure}
Even though the initial conditions (\ref{eq:Anonlinear}) involve only few gauge field components, the non-linear dynamics very efficiently includes the other components as well. Similar to Fig.~\ref{fig1}, we show in Fig.~\ref{fig:NO-A-evolution} the time evolution for all gauge field components. The three lines in each graph correspond to the different momenta 
$p_z = 0$, $p_z = 0.48 (g^2 \epsilon)^{1/4}$ and 
$p_z = 1.2 (g^2 \epsilon)^{1/4}$, respectively. The zero momentum mode 
oscillations of $A^2_x$ and $A^3_y$, expected from the linear analysis of Sec.~\ref{sec:linear}, are manifest as well as the primary growth in these components at non-zero momenta and in $A^2_x$ and $A^3_y$.\footnote{At early times the magnetic field is approximately given by $ B_z^1(t) \simeq A_x^2(t,0) A_y^3(t,0)$, because the
contribution of the derivative terms is small.} The simulations reveal growth for all other components such that most of the details about the initial conditions are lost rather quickly. All growth saturates at approximately 
$ t \simeq 40 \, (g^2\epsilon)^{1/4}$, which coincides with the time when the zero modes cease to oscillate. Before this time these oscillations are almost identical to the unperturbed case described by (\ref{eq:Asol}).

The initial configuration (\ref{eq:Blong}) pointing in the longitudinal direction is highly anisotropic and an important question in this context is the characteristic time for isotropization. For applications to hydrodynamic descriptions of heavy ion collisions the isotropization time of the stress-energy tensor is of particular relevance. The diagonal entries of the spatial components of the stress-energy tensor, or transverse and longitudinal 'pressure', are shown as a function of time in Fig.~\ref{fig-NO-T_ij}. One observes that for the macroscopic initial $B$-field configuration described by (\ref{eq:Anonlinear}) the stress-energy tensor shows oscillatory behavior. We find significant damping of these oscillations and approximate isotropization about the time when the exponential growth with characteristic rate $\sqrt{gB}$ stops. As we will see in Sec.~\ref{sec:iso}, this result can change dramatically if fluctuations are included in the initial conditions.

\section{Including initial fluctuations}
\label{sec:iso}

\subsection{Ensemble of coherent fields}
\label{sec:coherent}

So far, we considered initial configurations with a sharply defined macroscopic field amplitude $B$ described by (\ref{eq:Anonlinear}). An important step towards more realistic scenarios concerns the inclusion of fluctuations in the initial conditions. Here we choose non-zero initial values of the homogeneous $A_i^a$ fields randomly from a Gaussian distribution with zero mean and finite width. Then we perform an ensemble average over the initial conditions. Such a nonequilibrium ensemble is meant to describe a single system in the same sense as, e.g., equilibrium thermodynamics may be applied to a macroscopic system.
More precisely, we consider at initial time the homogeneous field configurations
\begin{equation}
 g A_x^2 = s_1,\qquad  g A_y^3 = s_2 \, ,
\label{eq:noisy-initial}
\end{equation}
where the real random numbers $s_1$ and $s_2$ fulfill 
\begin{equation}
 \langle s_1 \rangle = \langle s_2 \rangle = 0 \, , \quad  
\langle s_1^2 \rangle = \langle s_2^2 \rangle = \Delta^2\, , 
\label{eq:noisy-initial-delta}
\end{equation}
where $\langle \ldots \rangle$ denotes the ensemble average with given width $\Delta$. Again, all other $A_i^a$ are taken to vanish initially up to a small
amplitude noise seeding the instabilities. 

\begin{figure}
\begin{center}
 \epsfig{file=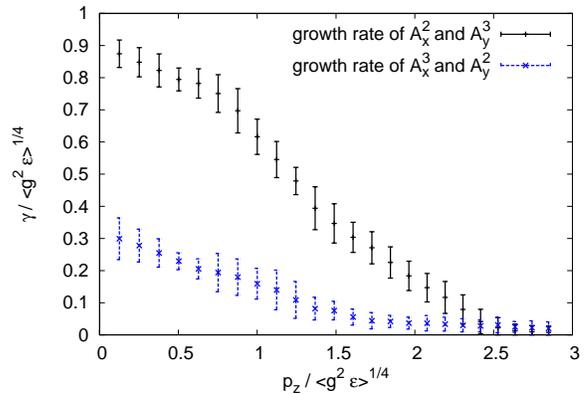, width=8cm}
 \caption{Growth rates for the fluctuations of gauge potentials, $\langle |A^2_x(t,p_z)|^2 \rangle^{1/2}$ etc., for the ensemble of 
coherent fields as described in the text. }
\label{fig-random-growthrates} 
\end{center}
\end{figure}
It turns out that growth rates in units of the average energy density
$\langle \epsilon \rangle$ are fairly independent of the value of the
width and the lattice size for the considered range of parameters.
Fig.~\ref{fig-random-growthrates} shows results for growth rates using
the range of width $\Delta a_s = 0.15-0.25 $ ensuring 
that the instability dynamics can be well resolved 
on the lattices we are using. We employ $64^3$ or
$128^3$ lattices for an ensemble built out of 1600 runs in total. The
vertical lines indicate the statistical errors. The ensemble
averages show a rather similar picture as for the initial condition
(\ref{eq:Anonlinear}) without fluctuations. For the latter case we saw
in Sec.~\ref{sec:linear} that $\delta A_y^3 - \delta A_x^2$ and
$\delta A_x^3 + \delta A_y^2$ exhibit primary growth rates as
displayed in Fig.~\ref{fig-NO-growth-rates}. Also for the ensemble
average the initially unstable modes turn out to be $A_x^2$, $A_y^3$, $A_y^2$ and
$A^3_x$.  However, two different growth rates are
observed from Fig.~\ref{fig-random-growthrates}. Here $\langle |A_x^2(t,p_z)|^2
\rangle^{1/2}$ and $\langle |A^3_y(t,p_z)|^2 \rangle^{1/2}$ exhibit a somewhat bigger
growth rate, while $\langle |A^3_x(t,p_z)|^2 \rangle^{1/2}$ and $\langle |A_y^2(t,p_z)|^2
\rangle^{1/2}$ have a smaller rate than for the simple initial condition
(\ref{eq:Anonlinear}).

\begin{figure}
\begin{center}
 \epsfig{file=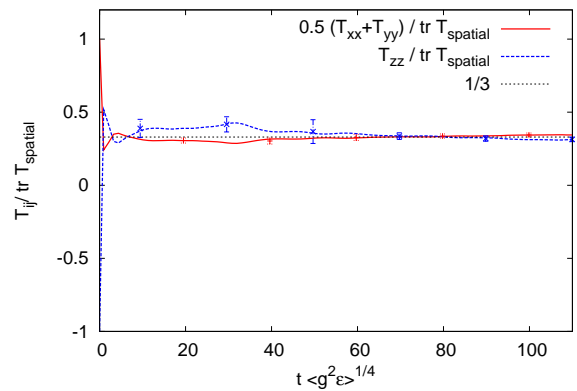, width=8cm}
 \caption{The same as in Fig.~\ref{fig-NO-T_ij}, however, including initial fluctuations described by the ensemble of coherent fields (\ref{eq:noisy-initial}). The vertical bars reflect the size of typical fluctuations on the smoothed curve we plot. The thin dashed line ($1/3$) corresponds to an isotropic $T_{ij}$.}
\label{fig-random-tmunu} 
\end{center}
\end{figure}
The inclusion of initial fluctuations has a significant impact on the behavior of the stress-energy tensor. In Fig.~\ref{fig-random-tmunu} we plot the diagonal entries of the spatial components of the stress-energy tensor as a function of time. Even though the initial configuration is highly anisotropic, one observes that transverse and longitudinal pressure very quickly approach the same value. In contrast to what has been observed in Sec.~\ref{sec:classical-statistical}, where approximate isotropy of pressure occured at the end of the period of exponential growth of fluctuations, it now happens on a much shorter time scale. This rapid isotropization is a dephasing phenomenon. For a sharply defined initial field amplitude $B$, which is realized with the initial condition (\ref{eq:Anonlinear}), the longitudinal and transverse pressure components oscillate with frequency $\sim \sqrt{g B}$ according to Fig.~\ref{fig-NO-T_ij}. The time average of the pressure over several oscillations was already isotropic in that case, but the coherent oscillations of the homogeneous
$A$ fields showed up in this quantity. In contrast, sampling many runs with different phases and initial field amplitudes leads very quickly to a practically constant value.
Rapid isotropization is an important ingredient for the application of hydrodynamic descriptions for heavy-ion collisions. However, in this case longitudinal expansion and increasing diluteness may make the system again become more anisotropic.   
We also emphasize that the observed dephasing phenomenon depends on the fact that the stress-energy tensor is dominated by the homogeneous modes for the considered initial
conditions. For the spatially inhomogeneous configurations in the context of 
color flux tubes, discussed in the following, this will no longer be the case.

\subsection{Modeling color flux tubes}
\label{sec:patches}

\begin{figure}
\begin{center}
\epsfig{file=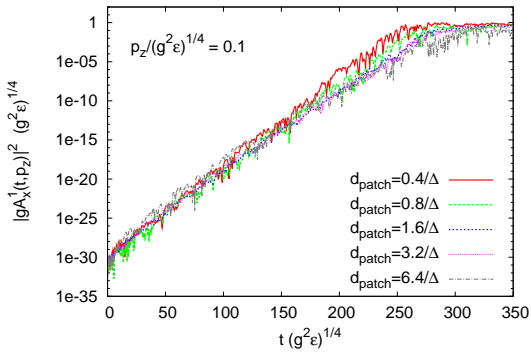, width=7.2cm, angle=0}
\caption{Time evolution of a low momentum mode of the gauge potential $A^1_y(t,p_z \simeq 0)$ for different sizes of correlated domains.}
\label{fig-patchsize}
\end{center}
\end{figure}
In this section, we also take into account that the fields may be correlated only over a limited transverse size while being homogeneous in longitudinal direction. 
To describe this, we divide the transverse plane
into square-shaped domains of equal size.  Each patch is filled with 
a coherent color magnetic field as in (\ref{eq:noisy-initial}), however, in each domain we use a different, randomly chosen color direction. 
The distribution of the random $A$-fields is a Gaussian with a 
given width $\Delta$. This corresponds to an inhomogeneous configuration, which is built from correlated domains of a characteristic transverse size. 
If the transverse size is taken to be $Q_s$  then this should model characteristic aspects of the dynamics of color flux tubes. The typical scale of the 
magnetic field inside a domain is $ g B_{\rm domain} \sim \Delta^2 $.

We consider domains ranging in size from 0.4 to 6 in units of the
characteristic spatial extent $1/\Delta$. As before, also a small
amplitude noise is added to all the fields to seed the instabilities. Similar
to the dynamics for the homogeneous initial conditions described
above, one observes also for the inhomogeneous configurations
approximately exponential growth of the spatial momentum modes of the
gauge potentials, $A^a_i(t,{\bf p})$. However, one observes primary growth for all color
and spatial components with approximately the same rate. 
In Fig.~\ref{fig-patchsize} the time evolution
for a low momentum mode\footnote{We use the first non-zero momentum mode on a
  $128^3$ lattice.} of the gauge potential, $A^1_x(t,p_z \simeq 0)$, is shown
as a function of time in units 
of the averaged energy density $\epsilon$
for systems with different initial domain
sizes. The data is obtained from an average over only eight runs, but since 
each run contains a significant number of domains with independent 
coherent fields it turns out that the growth rates are fairly stable. 

\begin{figure}
\begin{center}
\epsfig{file=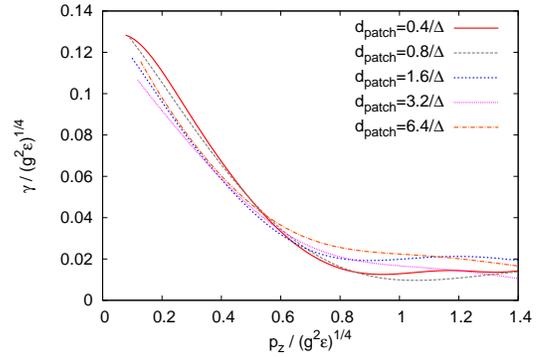, width=7.2cm, angle=0}
\caption{ The growth rate of a function of the momentum for various 
domain sizes.}
\label{fig-patchrate}
\end{center}
\end{figure}
In Fig.~\ref{fig-patchrate} the growth
rates are given as a function of momenta for different domain sizes. 
Comparing the results for the homogeneous and the
inhomogeneous initial configurations, one observes a similar 
momentum dependence, but the latter
show smaller growth rates by approximately a factor of
six to eight. It is interesting to note that one still might understand this difference in terms of the characteristic size of the space-averaged color-magnetic fields. This average is 
not $ B_{\rm domain}$ because different domains have independent fields,
which can average out. Instead, the characteristic size may be associated to the
average of the absolute value of the magnetic field.
 In the case of the 'flux tube' 
initial condition the space-average of the magnetic field is 
typically a factor of hundred smaller than in the homogeneous case 
in units of the energy density.
If the growth rates scale with the square root of the magnetic field, as is characteristic for a Nielsen-Olesen instability, one expects a
factor of ten smaller growth rates. However, this is only a lower estimate, 
since somewhat larger growth rates can occur if the spatial variations of 
the magnetic field are such that there are regions with 
larger magnetic fields which cancel in the average.

Remarkably, the growth rates measured
in units of the energy density of the system are similar to the growth
rates observed in Ref.~\cite{Berges:2007re}. In the latter study
inhomogeneous initial field configurations were randomly generated
from a Gaussian distribution. The oblate initial distribution for the
spatial Fourier modes $|A^a_i(t=0,{\bf p})|$ was characterized by a
transverse width $\Delta_T$ and a longitudinal width $\Delta_L$, where
$\Delta_L \ll \Delta_T$. This corresponds to an extreme anisotropy
with a $\delta(p_z)$-like distribution in longitudinal direction. As a
consequence, the approximately homogeneous field configurations in
longitudinal direction employed in Ref.~\cite{Berges:2007re} resemble
quite closely the correlated domains employed here if the transverse
width $\Delta_T$ is associated to the inverse domain size of our
configurations.

\section{Conclusions}
\label{sec:conclude}

In this paper we have studied out-of-equilibrium dynamics in $SU(2)$
pure gauge field theory. We presented a detailed 
analytic calculation of the time evolution in the linear regime starting 
from coherent field configurations. Generalizing the well-known Nielsen-Olesen instabilities for constant color-magnetic fields, in a first step we took into account temporal modulations. This leads to a remarkable coexistence of the original Nielsen-Olesen instability for time-averages together with a subdominant instability band because of the phenomenon of parametric resonance. The comparison to classical-statistical lattice gauge theory simulations showed remarkable agreement between analytics and numerics in the linear regime, while the lattice results show important nonlinear phenomena at later times such as secondary growth rates which are multiples of primary growth rates. 

This analysis provided the building blocks for the understanding of the dynamics of more realistic initial conditions including fluctuations. For the ensemble of coherent fields, which have zero mean color-magnetic field, again robust growth of the Nielsen-Olesen type is observed.
Remarkably, we find that for this ensemble isotropization of the stress-energy tensor happens on a much shorter time scale than the characteristic inverse growth rate of the instability. This rapid isotropization is a dephasing phenomenon. Rapid isotropization is an important ingredient for the application of hydrodynamic descriptions for heavy-ion collisions. However, in this case longitudinal expansion and increasing diluteness~\cite{Romatschke:2005pm,Fukushima:2011nq} can make the system again become more anisotropic. In particular, the observed dephasing phenomenon depends on the fact that the stress-energy tensor is dominated by the homogeneous modes for the considered initial
conditions. For spatially inhomogeneous configurations this is typically not the case.

To see this and in order to make the link to earlier work on plasma instabilities using inhomogeneous initial field configurations, we finally took into account that the fields should be correlated only over a limited transverse size. If the transverse size is taken to be $Q_s$ or $\sqrt{gB}$ then this should model characteristic aspects of the dynamics of color flux tubes. In this case we observed exponential growth for all components of the inhomogeneous gauge field configuration. The nature of the underlying instability still shows properties reminiscent of the Nielsen-Olesen--type. In particular, the results indicate that the characteristic dependence of the primary growth rate as a function of momentum still reaches its maximum at low momenta. In contrast, Weibel instabilities are expected to show a vanishing growth rate at zero momentum~\cite{Weibel}, which has also been emphasized in Ref.~\cite{Iwazaki:2008xi}. In view of these results it is interesting to observe that previous findings about plasma instabilities from classical-statistical lattice gauge theory~\cite{Romatschke:2005pm,Berges:2007re,Fukushima:2011nq} share important aspects of a Nielsen-Olesen instability.

In this work, we have not discussed the extraction of distribution functions and their possible interplay with time-dependent condensates after isotropization as suggested recently in Ref.~\cite{Blaizot:2011xf}. Distribution functions may, for instance, be derived from equal-time correlation functions in Coulomb gauge using the classical-statistical simulations employed in this work. This is deferred to a separate publication~\cite{preparation}.

\acknowledgments \noindent
The authors would like to thank Hiro Fujii for many inspiring discussions. They acknowledge the support by the Deutsche Forschungsgemeinschaft, the University of Heidelberg (FRONTIER), the Alliance Program of the Helmholtz Association (HA216/EMMI), by BMBF and MWFK Baden-W\"urttemberg (bwGRiD cluster).

\section*{Appendix}
\subsection{Linear regime including parametric resonance}
In Sec.~\ref{sec:linear} we discussed analytic solutions in the linear regime, which concentrated on the dominant exponentially growing modes at low momenta. However, the analysis neglects the phenomenon of parametric resonance in the presence of a periodically evolving background field. In this appendix we show that this phenomenon leads to a subdominant instability band at higher longitudinal momenta than the Nielsen-Olesen instability. Here we focus on the evolution of $\delta A_-(t,p)$ described by (\ref{eq:Ap}) of Sec.~\ref{sec:linear} as this shows the dominant primary instability. Plugging in the solution for the evolution of the background field (\ref{eq:Asol}), the evolution equation for vanishing transverse momentum reads
\begin{eqnarray}
\left[\partial_t^2+p^2-gB~\text{cn}^2\left(\sqrt{gB}~t,\frac{1}{2}\right) \right]\delta A_-(t,p)=0 ,
\label{eq:AppEVO}
\end{eqnarray}
where in this appendix we always denote the longitudinal momentum by $p$ to ease the notation. 
This closely resembles the Jacobian form of the Lam\'{e} equation. The crucial difference here is the negative sign in front of the oscillating term, which gives rise to the Nielsen-Olesen type instability discussed in Sec.~\ref{sec:linear}. Our strategy for solving the above evolution equation consists of approximating
\begin{eqnarray}
\text{cn}^2(x,m)=1-\text{cn}^2(x-K(m),m)+\mathcal{O}(m^2)
\label{eq:APPapprox}
\end{eqnarray}
for $m=1/2$ in (\ref{eq:AppEVO}) in order to reduce it to the well known Lam\'{e} equation, where again $K(m)$ denotes the complete elliptic integral of the first kind \cite{Elliptic}. While this approximation captures the important features of (\ref{eq:AppEVO}), it has the shortcoming that the average field strength $g\overline{B}$ according to Eq. (\ref{eq:averageB}) is overestimated by a factor of
\begin{eqnarray}
\frac{g\overline{B}}{g\overline{B}_{\text{app}}}=\frac{c}{1-c} \;, \quad c=\frac{ \Gamma(3/4)^2 }{ \Gamma(5/4) \Gamma(1/4)} \simeq 0.457 \;.
\label{eq:APPavgB}
\end{eqnarray}
Consequently, we expect the approximation to yield slightly enhanced growth rates that extent to somewhat higher momenta within about ten percent accuracy of the full numerical solution at early times, where the linear analysis is expected to hold. In order to obtain the solutions for $\delta A_-(t,p)$ for the above approximation, we recast the evolution equation to the Weierstrass form by use of the identity \cite{Boyanovski,Elliptic}
\begin{eqnarray}
\text{cn}^2\left(x,\frac{1}{2}\right)=-2\, \wp\left(x+iK\left(\frac{1}{2}\right)\right) \;,
\label{eq:APPwpcn}
\end{eqnarray}
where $\wp(x)$ is the Weierstrass elliptic function with roots $e_1=1/2$, $e_2=0$ and $e_3=-1/2$. The evolution equation in Weierstrass form then reads
\begin{eqnarray}
\left[\partial_t^2+p^2-gB-2gB\, \wp\left(\sqrt{gB} t-\tau)\right)\right]\delta A_-(t,p)=0 \;,
\nonumber\\
\label{eq:APPWSeq}
\end{eqnarray}
where $\tau=(1-i)K(1/2)$. By introducing the dimensionless time variable $\theta=\sqrt{gB}\, t$ and expressing the time independent contribution in (\ref{eq:APPWSeq}) in terms of the Weierstrass elliptic function according to
\begin{eqnarray}
\wp(z)=1-\frac{p^2}{gB} \, ,
\label{eq:APPzDef}
\end{eqnarray}
the fundamental solutions $U_{p}^{1}(\theta)$ and $U_{p}^{2}(\theta)$ to (\ref{eq:APPWSeq}) read \cite{ODE}
\begin{eqnarray}
U_{p}^{1/2}(\theta)=e^{\mp(\theta-\tau)\zeta(z)}\,\frac{\sigma(\theta-\tau \pm z)}{\sigma(\theta-\tau)} \, .
\end{eqnarray}
Here the superscript $1/2$ labels the respective fundamental solution.
The functions $\sigma(x)$ and $\zeta(x)$ represent the corresponding Weierstrass functions and the momentum dependence of the solutions is encoded in $z$ defined by (\ref{eq:APPzDef}).
\begin{figure}[t]
\begin{center}
\epsfig{file=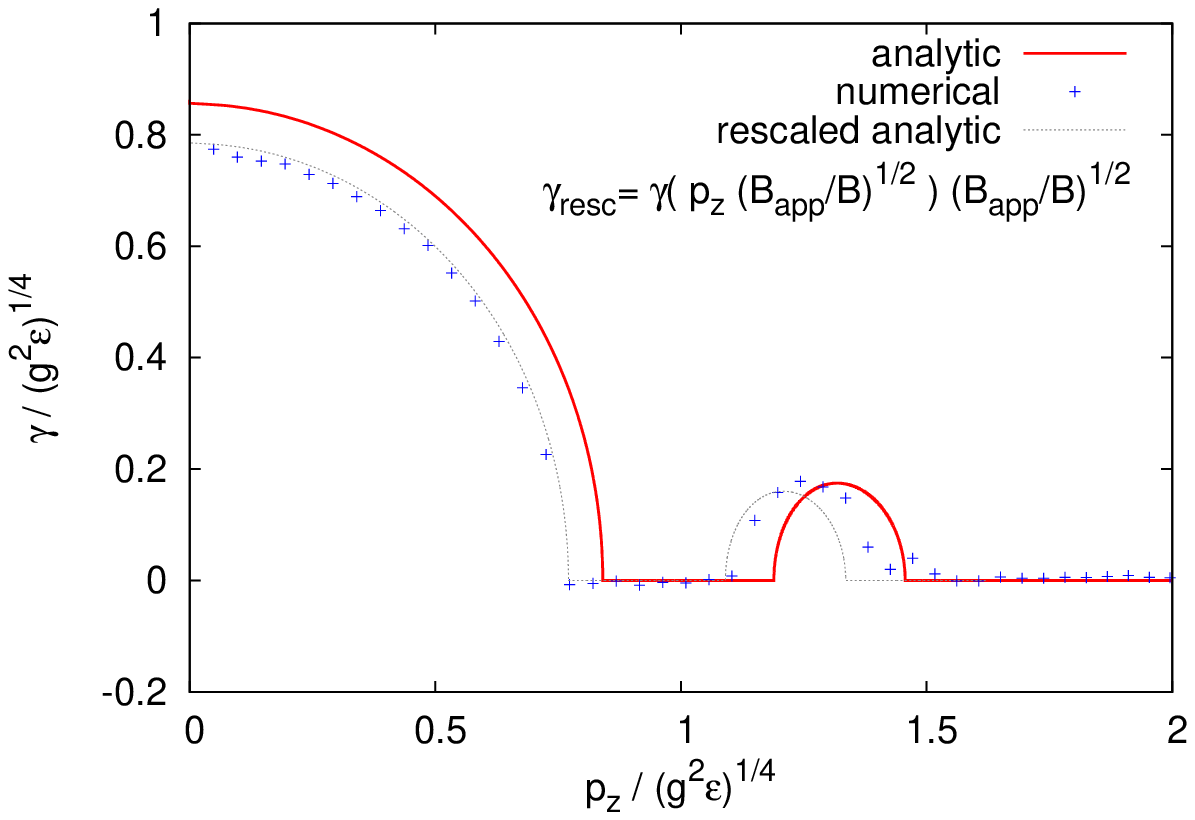, width=7.2cm, angle=0}
\epsfig{file=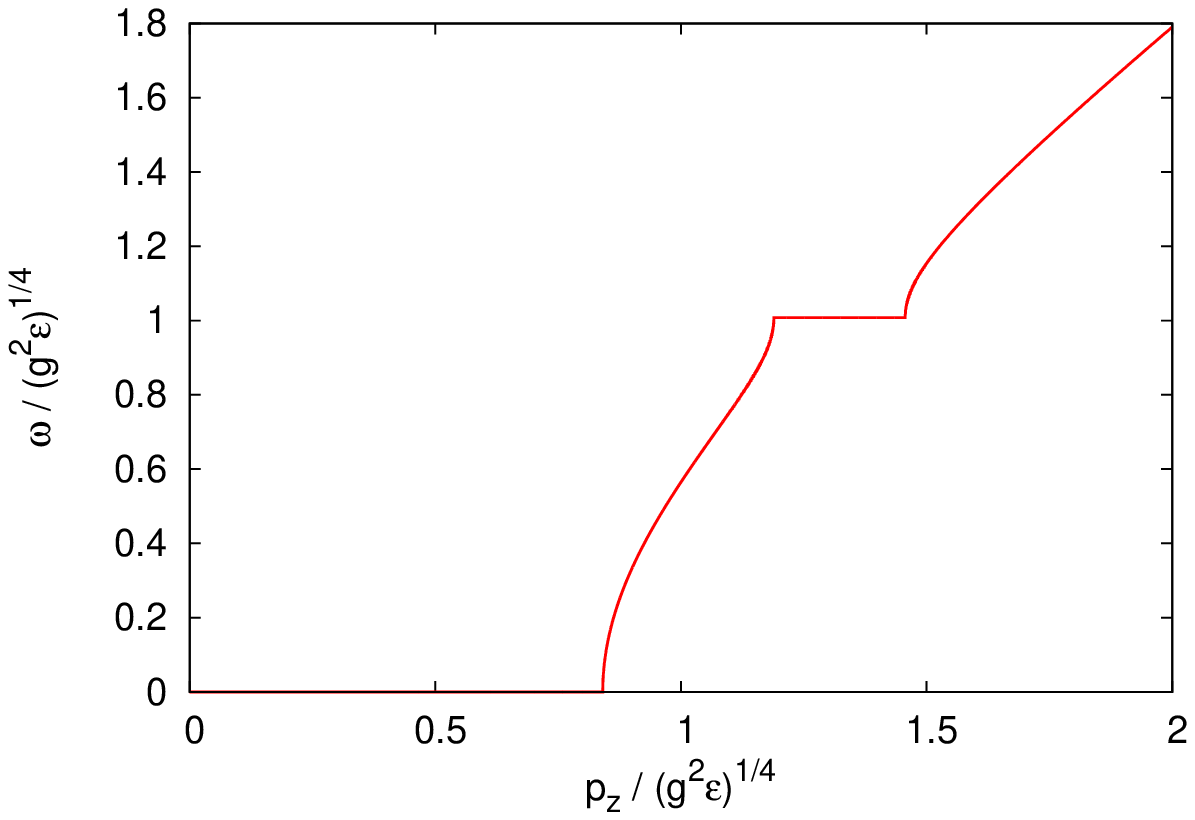, width=7.2cm, angle=0}
\caption{\label{fig-anares} Growth rates and oscillation frequencies in the linear regime.}
\end{center}
\end{figure}

In order to obtain the growth rate $\gamma(p)$ and the oscillation frequency $\omega(p)$ we proceed along the lines of Sec.~\ref{sec:linear} performing a Floquet analysis, i.e.
\begin{eqnarray}
U_{p}^{1/2}(\theta+2K(1/2))=e^{i\,C_-^{1/2}(p)}\, U_{p}^{1/2}(\theta) 
\end{eqnarray}
with the Floquet function $C_{-}^{1/2}(p)$ as in Sec.~\ref{sec:linear}. Using the quasi periodicity of the Weierstrass $\sigma$-function \cite{Boyanovski,Elliptic}
\begin{eqnarray}
\sigma(x+2K(1/2))=-\sigma(x)\exp[2(x+K(1/2))\zeta(K(1/2))] \nonumber \\
\end{eqnarray}
we find
\begin{eqnarray}
C^{1/2}_-(p)=\pm2i\left[K(1/2)\zeta(z)-z~\zeta(K(1/2))\right]\;.
\end{eqnarray}
The growth rate $\gamma(p)$ and oscillation frequency $\omega(p)$ are related to the real and imaginary parts of the floquet index by\footnote{We note that in this way the oscillation is only defined up to constants of $2\pi$.}
\begin{eqnarray}
\gamma(p)&=&\sqrt{gB}~\frac{|\text{Im}[C_-(p)]|}{2K(1/2)} \label{eq:APPgamma}\\
\omega(p)&=&\sqrt{gB}~\frac{|\text{Re}[C_-(p)]|}{2K(1/2)} \label{eq:APPomega}\;.
\end{eqnarray}
In order to evaluate the expressions in (\ref{eq:APPgamma}) and (\ref{eq:APPomega}) one needs to evaluate the mapping (\ref{eq:APPzDef}) defining $z$. Following the lines of \cite{Boyanovski} we find that there are four different regimes described by
\begin{eqnarray}
z&=&\beta \qquad \qquad \qquad~~ \text{for} \qquad \qquad \frac{p^2}{gB}<\frac{1}{2} \, , \label{eq:APPB1}\\
z&=&i\beta+K(1/2) \qquad \text{for} \qquad~ \frac{1}{2}<\frac{p^2}{gB}<1 \, , \label{eq:APPB2} \\
z&=&\beta+i K(1/2) \qquad \text{for} \qquad~ 1<\frac{p^2}{gB}<\frac{3}{2} \, , \label{eq:APPB3} \\
z&=&i\beta \qquad \qquad \qquad ~ \text{for} \qquad \qquad \frac{p^2}{gB}>\frac{3}{2} \, , \label{eq:APPB4}
\end{eqnarray}
where $\beta \in [0,K(1/2)]$ for all regimes. The different regimes correspond to two stable and two unstable bands. The lowest band corresponds to the Nielsen-Olesen instability, whereas the third band corresponds to the parametric resonance instability. Outside these momentum regions all modes are stable within the approximation using (\ref{eq:APPapprox}). 

The corresponding growth rates and oscillation frequencies for all bands are on display in Fig.~\ref{fig-anares}. The modes being subject to the Nielsen-Olesen instability show no oscillatory behavior according to this approximation. In contrast, modes which are amplified by the parametric resonance instability follow the oscillation of the macroscopic field $\bar{A}(t)$. The stable band in between interpolates between the two oscillation frequencies, while at higher momenta $p^2\gg gB$ the free field limit $\omega(p)=p$ is approached. Concerning the growth rates one observes that the Nielsen-Olesen instability exhibits the dominant growth rate when compared to parametric resonance. 
When comparing the analytic results for the growth rate to lattice data, we find that the analytic prediction is somewhat larger and extends to higher momenta. As mentioned above, this is a consequence of the approximation (\ref{eq:APPapprox}). However, one can accurately describe the lattice data by a trivial rescaling of the magnetic field amplitude $B$ in the analytical result by a factor of $\sqrt{c/(1-c)}$ as suggested by (\ref{eq:APPavgB}) to reproduce the correct average amplitude $\overline{B}$. 

\subsection{Parametric resonance band}
The above discussion provided implicit expressions for the growth rate as a function of momentum. To obtain explicit expressions one has to evaluate the variable $z$ introduced in (\ref{eq:APPzDef}) as a function of the momentum $p$. According to (\ref{eq:APPB1})-(\ref{eq:APPB4}) this has to be done separately for each band. In Sec.~\ref{sec:linear} we have already provided an explicit expression for the growth rate of the Nielsen-Olesen instability by an alternative approach. The same is desireable for the parametric resonance instability. To obtain such an expression we evaluate $z$ as a function of $p$, where for the parametric resonance band (\ref{eq:APPB3}) one has 
\begin{eqnarray}
z=\beta+iK(1/2)\;,
\end{eqnarray}
with
\begin{eqnarray} 
\beta(p)=\text{cn}^{-1}\left(\sqrt{2}~\sqrt{\frac{p^2}{gB}-1}~;~\frac{1}{2}\right)\;.
\label{eq:APPbetaPR}
\end{eqnarray}
The last expression was obtained by expressing the Weierstrass function $\wp(z)$ in (\ref{eq:APPzDef}) in terms of Jacobi elliptic functions according to (\ref{eq:APPwpcn}). Though (\ref{eq:APPbetaPR}) along with (\ref{eq:APPgamma}) already provides an explicit expression for the growth rate, it proves insightful to apply further approximations. In a first step we expand the Weierstrass zeta function as
\begin{eqnarray}
\zeta(x)&=&\frac{\eta x}{\omega}+\frac{\pi}{2\omega}\cot\left(\frac{\pi x}{2\omega}\right)+\frac{2\pi}{\omega}\sum_{k=1}^{\infty}\frac{q^{2k}}{1-q^{2k}}\sin\left(\frac{k\pi x}{\omega}\right) \nonumber \\ \\
\eta&=&\zeta(\omega)=\frac{\pi^2}{12\omega}+\mathcal{O}(q^2)
\end{eqnarray}
where $q=e^{-\pi}$ and $\omega=K(1/2)$. Without further approximations the growth rate $\gamma(p)$ is then given by
\begin{eqnarray}
\gamma(p)&=&\sqrt{gB}~\Bigg|\text{Re}\left[\frac{\pi}{2\omega}\cot\left(\frac{\pi z}{2\omega}\right)\frac{}{}\right. \nonumber \\ &+&\left.\frac{2\pi}{\omega}\sum_{k=1}^{\infty}\frac{q^{2k}}{1-q^{2k}}\sin\left(\frac{k\pi z}{\omega}\right)\right]\Bigg| 
\end{eqnarray}
The real part can then be expanded in a $q$-Series by seperately expanding the cotangent and the series of sine functions. As $q\simeq0.043$ is a reasonably small expansion parameter we will keep the leading terms only. For the cotangent the real part is given by
\begin{eqnarray}
\text{Re}\left[\cot\left(\frac{\pi}{2\omega}\beta+i\frac{\pi}{2}\right)\right]&=&\frac{\sin(\frac{\pi}{\omega}\beta)}{\cosh(\pi)-\cos(\frac{\pi}{\omega}\beta)} \nonumber \\
&=&2q\sin\left(\frac{\pi}{\omega}\beta\right)+\mathcal{O}(q^2)\;,
\end{eqnarray}
and similarly for the series of sine functions one finds
\begin{eqnarray}
4\text{Re}\left[\sum_{k=1}^{\infty}\frac{q^{2k}}{1-q^{2k}}\sin\left(\frac{\pi}{\omega}k\beta+ik\pi)\right)\right]&=& \nonumber \\
2\sum_{k=1}^{\infty}\frac{q^{k}+q^{3k}}{1-q^{2k}}\sin\left(\frac{\pi}{\omega}k\beta\right)&=&\nonumber \\
2q\sin\left(\frac{\pi}{\omega}\beta\right)+\mathcal{O}(q^2)\;.
\end{eqnarray}
As these expressions involve trigonometric functions of $\beta(p)$ a simple expression for the momentum dependent growth rate $\gamma(p)$ in the parametric resonance regime can be obtained by expanding the inverse of the Jacobi cosine in (\ref{eq:APPbetaPR}) in terms of inverse trigonometric functions according to 
\begin{eqnarray}
\text{cn}^{-1}(x,m)=\frac{2\omega}{\pi}\cos^{-1}(x)+\mathcal{O}(m) \;.
\end{eqnarray}
By use of double angle formulas we obtain as the final result
\begin{eqnarray}
\gamma(p)\simeq \sqrt{gB}~e^{-\pi}~\frac{8\pi}{K(1/2)}~\sqrt{\frac{p^2}{gB}-1}~\sqrt{\frac{3}{2}-\frac{p^2}{gB}}
\;\end{eqnarray}
for the parametric resonance band where $1<p^2/(gB)<3/2$. The maximal growth rate $\gamma_0$ is realized at the momentum $p \simeq\sqrt{5/4} \sqrt{gB}$. Numerically one finds $\gamma_0\simeq0.146 \sqrt{gB}$ which is significantly smaller than the growth rates associated with the Nielsen-Olesen instability.


\begin{thebibliography}{10}

%\cite{Aarts:2001yn}
\bibitem{Aarts:2001yn}
  G.~Aarts, J.~Berges,
  %``Classical aspects of quantum fields far from equilibrium,''
  Phys.\ Rev.\ Lett.\  {\bf 88 } (2002)  041603.
  %[hep-ph/0107129].

%\cite{Arrizabalaga:2004iw}
\bibitem{Arrizabalaga:2004iw}
  A.~Arrizabalaga, J.~Smit, A.~Tranberg,
  %``Tachyonic preheating using 2PI-1/N dynamics and the classical approximation,''
  JHEP {\bf 0410 } (2004)  017.
  %[hep-ph/0409177].

%\cite{Berges:2008wm}
\bibitem{Berges:2008wm}
  J.~Berges, A.~Rothkopf, J.~Schmidt,
  %``Non-thermal fixed points: effective weak-coupling for strongly correlated
  %systems far from equilibrium,''
  Phys.\ Rev.\ Lett.\  {\bf 101} (2008) 041603.
  %[arXiv:0803.0131 [hep-ph]].
  %%CITATION = PRLTA,101,041603;%%

%\cite{Berges:2010zv}
\bibitem{Berges:2010zv}
  J.~Berges, D.~Gelfand, J.~Pruschke,
  %``Quantum theory of fermion preheating,''
  Phys.\ Rev.\ Lett.\  {\bf 107} (2011) 061301.
  %[arXiv:1012.4632 [hep-ph]].
  %%CITATION = PRLTA,107,061301;%%

%\cite{Romatschke:2005pm}
\bibitem{Romatschke:2005pm}
  P.~Romatschke, R.~Venugopalan,
  %``Collective non-Abelian instabilities in a melting color glass condensate,''
  Phys.\ Rev.\ Lett.\  {\bf 96 } (2006)  062302.
  %[hep-ph/0510121].
%\cite{Romatschke:2006nk}
%\bibitem{Romatschke:2006nk}
  P.~Romatschke, R.~Venugopalan,
  %``The Unstable Glasma,''
  Phys.\ Rev.\  {\bf D74} (2006) 045011.
  %[hep-ph/0605045].

%\cite{Berges:2007re}
\bibitem{Berges:2007re}
  J.~Berges, S.~Scheffler, D.~Sexty,
  %``Bottom-up isotropization in classical-statistical lattice gauge theory,''
  Phys.\ Rev.\  {\bf D77 } (2008)  034504.
  %[arXiv:0712.3514 [hep-ph]].
%\cite{Berges:2008zt}
%\bibitem{Berges:2008zt}
  J.~Berges, D.~Gelfand, S.~Scheffler, D.~Sexty,
  %``Simulating plasma instabilities in SU(3) gauge theory,''
  Phys.\ Lett.\  {\bf B677 } (2009)  210.
  %[arXiv:0812.3859 [hep-ph]].
%\cite{Berges:2008mr}
%\bibitem{Berges:2008mr}
  J.~Berges, S.~Scheffler, D.~Sexty,
  %``Turbulence in nonabelian gauge theory,''
  Phys.\ Lett.\  {\bf B681 } (2009)  362.
  %[arXiv:0811.4293 [hep-ph]].

%\cite{Kunihiro:2010tg}
\bibitem{Kunihiro:2010tg}
  T.~Kunihiro, B.~Muller, A.~Ohnishi, A.~Schafer, T.~T.~Takahashi, A.~Yamamoto,
  %``Chaotic behavior in classical Yang-Mills dynamics,''
  Phys.\ Rev.\  {\bf D82 } (2010) 114015.
  %[arXiv:1008.1156 [hep-ph]].

%\cite{Fukushima:2011nq}
\bibitem{Fukushima:2011nq}
  K.~Fukushima, F.~Gelis,
  %``The evolving Glasma,''
  arXiv:1106.1396 [hep-ph].

%\cite{Gelis:2010nm}
\bibitem{Gelis:2010nm}
  For a recent review see F.~Gelis, E.~Iancu, J.~Jalilian-Marian, R.~Venugopalan,
  %``The Color Glass Condensate,''
  Ann.\ Rev.\ Nucl.\ Part.\ Sci.\  {\bf 60 } (2010)  463.
  %[arXiv:1002.0333 [hep-ph]].

%\cite{Iwazaki:2008xi}
\bibitem{Iwazaki:2008xi}
  A.~Iwazaki,
  %``Decay of Color Gauge Fields in Heavy Ion Collisions and Nielsen-Olesen Instability,''
  Prog.\ Theor.\ Phys.\  {\bf 121 } (2009)  809.
  %[arXiv:0803.0188 [hep-ph]].
%\cite{Fujii:2008dd}
%\bibitem{Fujii:2008dd}
  H.~Fujii, K.~Itakura,
  %``Expanding color flux tubes and instabilities,''
  Nucl.\ Phys.\  {\bf A809 } (2008)  88.
  %[arXiv:0803.0410 [hep-ph]].
%\cite{Fujii:2009kb}
%\bibitem{Fujii:2009kb}
  H.~Fujii, K.~Itakura, A.~Iwazaki,
  %``Instabilities in non-expanding glasma,''
  Nucl.\ Phys.\  {\bf A828 } (2009)  178.
  %[arXiv:0903.2930 [hep-ph]].

\bibitem{NO} N.K.~Nielsen, P.~Olesen, Nucl.\ Phys.\ {\bf B144} (1978) 376; Phys.\ Lett.\ {\bf B79} (1978) 304. S.J.~Chang, N.~Weiss, Phys.\ Rev.\ {\bf D20} (1979) 869.

\bibitem{Weibel} P.~Romatschke, M.~Strickland, Phys.\ Rev.\ {\bf D68} (2003) 036004; P.~Arnold, J.~Lenaghan,
G.D.~Moore, JHEP {\bf 08} (2003) 002.

%\cite{Berges:2008zt}
\bibitem{Berges:2008zt}
  J.~Berges, D.~Gelfand, S.~Scheffler, D.~Sexty,
  %``Simulating plasma instabilities in SU(3) gauge theory,''
  Phys.\ Lett.\  {\bf B677 } (2009)  210.
  %[arXiv:0812.3859 [hep-ph]].

%\cite{Ipp:2010uy}
\bibitem{Ipp:2010uy}
  A.~Ipp, A.~Rebhan, M.~Strickland,
  %``Non-Abelian plasma instabilities: SU(3) vs. SU(2),''
  arXiv:1012.0298 [hep-ph].

\bibitem{chaos}
S.G.~Matinyan, G.K.~Savvidi, N.G.~Ter-Arutyunyan-Savvidi, Zh.\ Eksp.\ Teor.\ Fiz.\ {\bf 80} (1981) 830.

%\cite{Kofman:1994rk}
\bibitem{Kofman:1994rk}
  L.~Kofman, A.~D.~Linde and A.~A.~Starobinsky,
  %``Reheating after inflation,''  
  Phys.\ Rev.\ Lett.\  {\bf 73} (1994) 3195.%  [hep-th/9405187].  %%CITATION = HEP-TH/9405187;%%

%\cite{Tudron:1980gq}
\bibitem{Tudron:1980gq}
  T.~N.~Tudron,
  %``Instability Of Constant Yang-mills Fields Generated By Constant Gauge Potentials,''
  Phys.\ Rev.\  {\bf D22 } (1980)  2566.
  
%\cite{Berges:2002cz}
\bibitem{Berges:2002cz}
  J.~Berges, J.~Serreau,
  %``Parametric resonance in quantum field theory,''
  Phys.\ Rev.\ Lett.\  {\bf 91 } (2003)  111601.

\bibitem{Elliptic}
M.~Abramowitz, I.A.~Stegun, eds.\ (1965), {\em Handbook of Mathematical Functions with Formulas, Graphs, and Mathematical Tables}, New York: Dover, pp. 587.

%\cite{Kovner:1995ja}
\bibitem{Kovner:1995ja}
  A.~Kovner, L.~D.~McLerran, H.~Weigert,
  %``Gluon production from nonAbelian Weizsacker-Williams fields in nucleus-nucleus collisions,''
  Phys.\ Rev.\  {\bf D52 } (1995)  6231.
  %[hep-ph/9502289].
%\cite{Krasnitz:1999wc}
%\bibitem{Krasnitz:1999wc}
  A.~Krasnitz, R.~Venugopalan,
  %``The Initial energy density of gluons produced in very high-energy nuclear collisions,''
  Phys.\ Rev.\ Lett.\  {\bf 84 } (2000)  4309.
  %[hep-ph/9909203].
%\cite{Lappi:2006fp}
%\bibitem{Lappi:2006fp}
  T.~Lappi, L.~McLerran,
  %``Some features of the glasma,''
  Nucl.\ Phys.\  {\bf A772 } (2006)  200.
  %[hep-ph/0602189].

\bibitem{Boyanovski}
D.~Boyanovsky, H.J.~de~Vega, R.~Holman, J.F.J.~Salgado,~arXiv:9608205v2~[hep-ph].

\bibitem{ODE}
E.L.~Ince, {\it Ordinary Differential Equations}, Dover 1944.

%\cite{Blaizot:2011xf}
\bibitem{Blaizot:2011xf}
  J.~-P.~Blaizot, F.~Gelis, J.~Liao, L.~McLerran, R.~Venugopalan,
  %``Bose--Einstein Condensation and Thermalization of the Quark Gluon Plasma,''  
  Nucl.\ Phys.\ A {\bf 873} (2012) 68. %  [arXiv:1107.5296 [hep-ph]].  %%CITATION = ARXIV:1107.5296;%%

\bibitem{preparation} J.~Berges, S.~Schlichting, D.~Sexty, to be published.

\end{thebibliography}
\end{document}